# Astro2020 Science White Paper

# Solar system Deep Time-Surveys of atmospheres, surfaces, and rings

**Thematic Areas:** ☒ Planetary Systems  ☐ Star and Planet Formation
☐ Formation and Evolution of Compact Objects  ☐ Cosmology and Fundamental Physics
☐ Stars and Stellar Evolution  ☐ Resolved Stellar Populations and their Environments
☐ Galaxy Evolution  ☐ Multi-Messenger Astronomy and Astrophysics

**Principal Author:**
Name: Michael H. Wong
Institution: UC Berkeley
Email: mikewong@astro.berkeley.edu
Phone: 510-224-3411

**Co-authors:** (names and institutions)
Richard Cartwright (SETI Institute)
Kunio Sayanagi (Hampton University)
Matthew Tiscareno (SETI Institute)
Glenn Orton (JPL)
James Sinclair (JPL)
Michael Lucas (UT Knoxville)
Bryan Holler (STScI)
Angel Otarola (TMT)
Katherine de Kleer (Caltech)
Patrick Fry (Univ. Wisconsin)
Statia H. Luszcz-Cook (AMNH)
Nancy Chanover (NMSU)
Thomas Greathouse (SwRI)
Rohini Giles (SwRI)
David Trilling (Northern Arizona Univeristy)
Noemi Pinilla-Alonso (Florida Space Institute, UCF)
Eric Gaidos (University of Hawaii)
Stephanie Milam (NASA GSFC)
Amy Simon (NASA GSFC)
Conor Nixon (NASA GSFC)
Máté Ádámkovics (Clemson Univ.)
Amanda Hendrix (PSI)

**Abstract:**
Imaging and resolved spectroscopy reveal varying environmental conditions in our dynamic solar system. Many key advances have focused on how these conditions change over time. Observatory-level commitments to conduct annual observations of solar system bodies would establish a long-term legacy chronicling the evolution of dynamic planetary atmospheres, surfaces, and rings. Science investigations will use these temporal datasets to address potential biosignatures, circulation and evolution of atmospheres from the edge of the habitable zone to the ice giants, orbital dynamics and planetary seismology with ring systems, exchange between components in the planetary system, and the migration and processing of volatiles on icy bodies, including Ocean Worlds. The common factor among these diverse investigations is the need for a very long campaign duration, and temporal sampling at an annual cadence.

**Background**

In the realm of astronomy, new discoveries are often enabled by the opening of new domains in terms of angular resolution, sensitivity, spectral bandpass, or spectral resolution. In much of modern planetary science, an equally important dimension for new discoveries is the time domain. For example, outer-planet atmospheres evolve on timescales from minutes (impacts), days (plumes and storms), weeks to months (evolution of zonal bands), to years (seasonal changes, impact aftermaths, upper atmospheric composition and structure, and evolution of large vortices). Lake distributions may shift on Titan, cloud activity varies on the ice giants, and volatiles migrate over the surfaces of Triton and the dwarf planets.

The Outer Planet Atmospheres Legacy (OPAL) program (Simon et al. 2015) may be the first example of a major astronomical observatory (HST) committing to a solar system Deep Time-Survey with an annual cadence. The long-term nature of the program has established a baseline of consistent data with yearly sampling since 2014, with productive science output (Wong 2017). OPAL contains specific elements making use of unique capabilities of HST: consecutive global maps for wind tracking (spanning 20–40 hours of planetary rotation) and UV imaging. Similarly, new Deep Time-Surveys at advanced future observatories would rely on the unique capabilities of these facilities to address science questions in the dynamic solar system.

**Overview of Deep Time-Surveys**

This white paper stems from a 2018 Key Science Proposal to the US Extremely Large Telescope (ELT) program. ELTs will carry major advantages in terms of angular resolution, advanced instrumentation, and light-gathering power. Other observatories will have their own specific advantages, in terms of wavelength range and observation timing. We envision a collective of solar system Deep Time-Surveys operating independently at each participating observatory, with each survey consisting of a diverse collection of science goals, united by a common temporal requirement: regular sampling on an annual timescale, continued over a decades-long duration that is particularly suited to the seasonal intervals in the outer solar system. With director-level commitments to Deep Time-Surveys, participation among multiple observatories will enable a wealth of multiwavelength studies of evolution in planetary atmospheres, surfaces, and rings. Conducting observations of bright solar system objects at morning twilight may reduce the impact on other ground-based programs (Otarola et al. 2015a).

For the US ELT program, we traced the path from high-level science goals to instrument and temporal requirements (Fig. 4). Similar, targeted objectives could be refined for other facilities like advanced millimeter/ microwave observatories or next-generation space observatories.

**Science Goals**

Science goals of this program are closely tied to Priority Questions from the last planetary decadal survey (*Visions and Voyages*), as indicated by numbers in column 1 of Fig. 4. We briefly describe individual science goals here:

*Search for life on Mars:*
Methane in the atmosphere of Mars has been detected from ground-based telescopes and orbiting and landed spacecraft; it could have exogenic, geologic, or even astrobiological origins. The observed concentration of $CH_4$ is highly variable with space and time, with transient pulses and hemispheric differences (Formisano et al. 2004, Mumma et al. 2009, Krasnopolsky et al. 2004,

Webster et al. 2015). Potential sinks for methane are also seasonally variable, so monitoring on timescales of Earth months to a year will help us understand the dynamics and distinguish between seasonal and non-seasonal variability (Encrenaz et al. 2004, Webster et al. 2018).

***Circulation and evolution of atmospheres on the edge of the habitable zone:***
Non-LTE emission from CO and $CO_2$ in the atmospheres of Mars and Venus constrains circulation in the mesosphere (Lellouch et al. 2000, Drossart et al. 2007), contributing to our understanding of the evolution of atmospheres toward states too hot or too cold for liquid surface water.

***Volatile cycles and seasonal variation on Titan:***
A resolved spectroscopic campaign would increase our understanding of variable cloud activity (e.g. Roe et al. 2008), as well as the $CH_4$ "hydrological" cycle (e.g. Teanby et al. 2008, Adamkovics et al. 2016), which includes effects such as evaporation from lakes (e.g., Brown et al., 2009; Turtle et al., 2009). Cloud activity patterns and trends in tropospheric $CH_4$ concentrations provide valuable tests for circulation models (Tokano 2014, Mitchell 2012), none of which match all the observations. A long-term campaign of Titan observations sensitive to atmospheric composition and clouds would provide continuity between the extensive observations of Titan by NASA's Cassini mission, and future missions such as the Dragonfly lander (Turtle et al. 2018).

***Heat transport, circulation, climate, and space weather of giant planets:***
Changes in coloration and reflectivity have been observed on regional and global scales in all the giant planets (Hammel and Lockwood 2007, Karkoschka 2011, Fletcher 2017), but the drivers for this evolution are not well understood, particularly since timescales are variable on Jupiter, and seasonal timescales extend for decades on the ice giants. Long-term observations will constrain the relevant timescales and periodicities (Fouchet et al. 2008), and observatories with high-resolution optical imaging capability will even enable cloud color measurements to characterize changes in aerosol properties at fine spatial scales (Simon et al. 2006, Wong et al. 2011). High-resolution imaging sequences will also be able to measure 2D flow on short timescales (Fig. 1), while compositional variation (diagnostic of vertical flow) can be derived by observations in the thermal infrared wavelength regime (Fig. 2) or at millimeter/microwave wavelengths. At higher altitudes, giant planet atmospheres interact with magnetospheric charged particles, ring rain, interplanetary dust, and impacting comets/asteroids, producing compositional spectral signatures that evolve over many years (Sinclair et al. 2017, O'Donoghue et al. 2017), as well as UV and IR ($H_3^+$) emission vary over shorter timescales (Grodent et al. 2018, Johnson et al. 2018). Do auroral spots vary with volcanic activity or water plumes, from Io, Europa, or Enceladus?

***Evolution of planetary rings:***
Planetary rings serve as accessible natural laboratories for disk processes, as clues to the origin and evolution of planetary systems, and as shapers and detectors of their planetary environments (e.g., Tiscareno and Murray 2018). Long-term or seasonal phenomena within the rings and small moons of the outer planets require time-domain observations (a cadence of one year would be ideal) for characterization. These include the ring arcs of Neptune, apparently Myr-unstable inner moons of Uranus (French and Showalter 2012), spokes and propellers in Saturn's rings, arcs in the rings of Jupiter, and moons with poorly understood orbital variations such as Daphnis and Prometheus. Evolving wave structures within planetary rings, potentially resolvable with ELTs and future space-based optical telescopes, record the impact history in the Jupiter (Showalter et al. 2011) and Saturn (Hedman et al. 2015) systems, as well as internal oscillations that constrain the deep structure of gas giants (Marley and Porco 1993, Mankovich et al. 2019).

*Understand the evolution of surfaces, including Ocean Worlds:*

High spatial resolution optical imaging could reveal new plume deposits on volcanically-active Io (Spencer et al. 1997). Large giant planet satellites and dwarf planets in the Kuiper Belt experience continual modification of their surfaces and atmospheres. Modification processes include: (1) bombardment by UV photons and energetic charged particles, (2) bombardment by heliocentric and planetocentric micrometeorites, (3) seasonal migration of volatiles, and (4) geologic activity (cryovolcanism, impact events, mass wasting, etc). Outer planet satellites and small bodies exhibit a dazzling array of processes relevant to the origin and evolution of the solar system, as well as the potential for extraterrestrial life. For example, magnetospheric irradiation produces disequilibrium chemistry (Chyba 2000) that could provide energy for life if exchange with subsurface oceans is active in the jovian system; E-ring material from cryovolcanic Enceladus accumulates on the surface of the Saturnian moons Mimas, Enceladus, Tethys, Dione, and Rhea (Hendrix et al. 2018); in the uranian system, $CO_2$ ice migrates due to varying insolation and solar heating (Grundy et al. 2006, Sori et al. 2017), allowing the surface distribution of dark, potentially carbonaceous material to be studied (Cartwright et al. 2018); the variation of frosts on Triton and KBOs may be related to transitions between global and local atmospheric states (Fig. 3, Holler et al. 2016, Cruikshank et al. 1993, Trafton 1984); there may be recent geologic activity, including possible glaciation and cryovolcanic structures on Pluto (Umurham et al., 2017; Moore et al., 2016, 2017); and aerosols generated via UV photolysis "rain out" onto Pluto's surface, incorporating into existing geologic provinces like Chulthu Macula (Gao et al. 2017, Grundy et al. 2018).

## Conclusion

Long-baseline campaign of solar system observations will enable a wide range of science, unified by a common need for annual/semiannual/biannual observations. The overarching motivation for these diverse studies fully spans the range of Decadal Survey topics of planetary origins, planetary evolution, interacting systems, and the search for life. Annual observing programs at multiple great observatories will produce an unprecedented collective dataset enabling multiwavelength studies of evolving planetary atmospheres, surfaces, and rings.

## Figures

**Neptune**

TMT 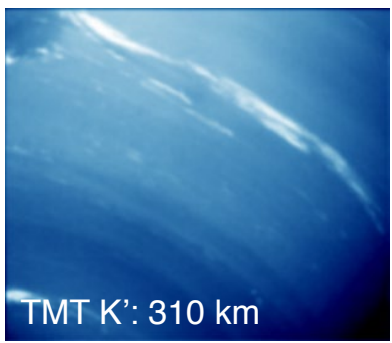  Voyager 2 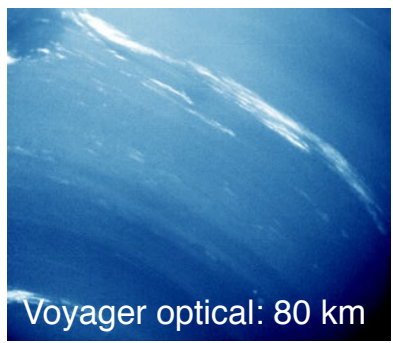

TMT K': 310 km  Voyager optical: 80 km

**Figure 1.** ELTs will achieve spatial resolution at the ice giants approaching what was achieved by the only spacecraft mission to visit them (Voyager 2). The image at left has been blurred to TMT resolution (at K band) to show that fine scale clouds will still be detected; methane clouds such as the white features seen by Voyager have even higher contrast in the IR. From Otarola et al. (2015b).

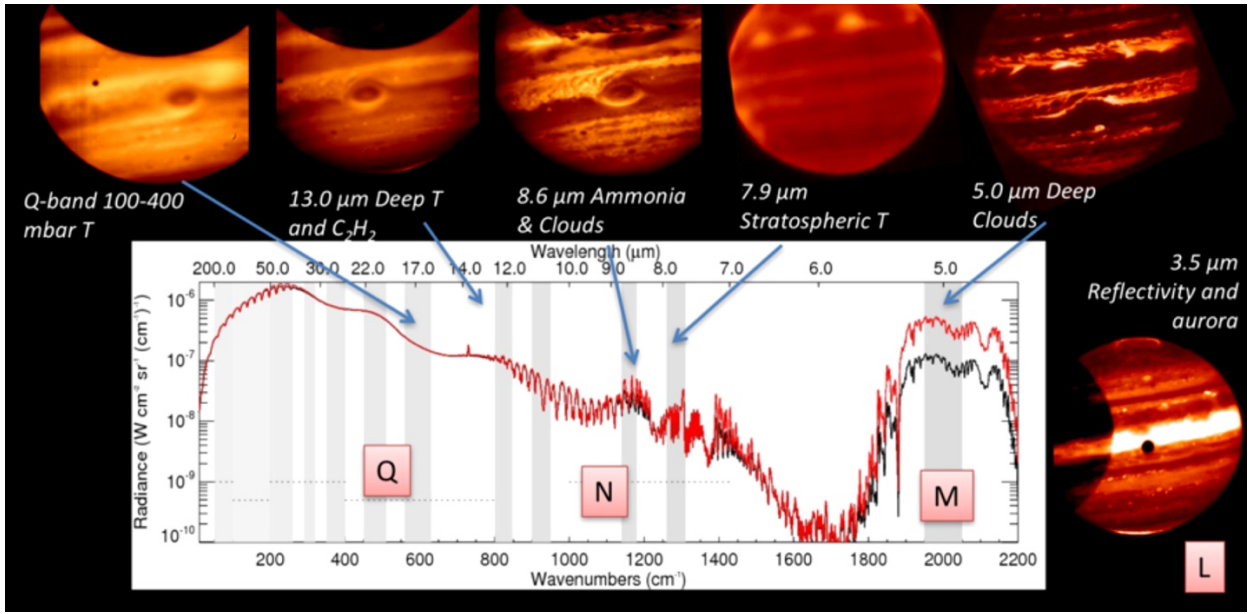

**Figure 2.** Time-resolved measurements of thermal emission reveal the evolution of temperatures and composition at multiple layers of the atmosphere. Observations at mid-IR wavelengths (shown here for Jupiter), as well as millimeter and radio wavelengths, are sensitive to atmospheric properties that trace chemical and dynamical processes.

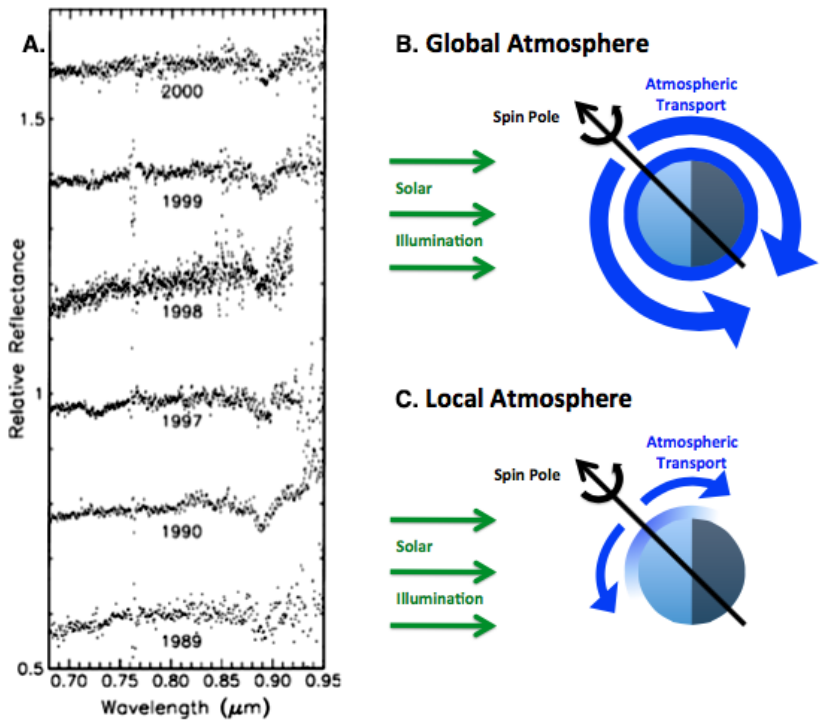

**Figure 3A.** Short-timescale spectral changes in Triton's reflectance spectrum suggest non-seasonal transport of $N_2$/CO/$CO_2$ frosts, which act to suppress spectral signatures of reddish material or $CH_4$ ice. Figure from Hicks and Hicks and Burratti (2004). **Figure 3B. and 3C.** The atmosphere of Eris may transition between a global state that can transport volatiles over the entire surface, and a local state where transport is limited to regions close to the warmest surface areas. Figure from Hofgartner et al. (2018).

| Science Goals | Science Objectives | Observables | Instrument requirements | Cadence | | | |
|---|---|---|---|---|---|---|---|
| | | | | Visits per year | Mosaic span | Mosaic step duration (hr) | Mosaic steps (N) |
| Search for life on Mars. [5,6] | Quantify CH4 on Mars: spatial and temporal variability | High resolution, spatially-resolved, synoptic spectra of Mars | NIR spectra at R~100,000. AO spatial resolution. | 2 | 25 hr | 0.5 | 6 |
| Circulation and evolution of atmospheres on the edge of the habitable zone. [5,6,9,10] | Trace mesosphere circulation | Spatially-resolved NIR and MIR spectra of Venus nightside and Mars dayside | TBD | TBD | TBD | TBD | TBD |
| Understand volatile cycles and seasonal variation on Titan, and similar processes in the solar system. [4,10] | Determine the spatial and temporal variation of haze layers, clouds, CH4 humidity, and surface moisture | Imaging and spatially-resolved spectroscopy of clouds and hazes on Titan | Diffraction limited imaging. Spectroscopy in 0.8–2.5 µm range at R~2000. | 1 | 16 d | 0.5 | 4 |
| Understand the heat transport, circulation, climate, and space weather of giant planets, and the relation to exoplanets and Earth. [7,9,10] | Determine the spatial distribution and vertical structure of cloud features on Uranus/Neptune | Spatially-resolved spectroscopy of clouds and hazes | Diffraction limited spectroscopy in 0.8–2.5 µm range at R~2000. | 2 | 15 h | 1 | 4 |
| | Measure 3D wind flow in all four giant planets | High-resolution image sequences for cloud tracking at multiple pressure-levels | Diffraction-limited imaging in NIR, range of ~5 narrow-band filters sampling different gas opacities. | 4 | 20-30 h | 0.5 | 16 |
| | Understand temporal variation of particle properties, composition, and temperature in zonal bands in all four giant planets | High-resolution images with global coverage | Diffraction-limited imaging in visible, NIR, and MIR. | 4 | 10-15 h | 0.5 | 6 |
| | Understand interactions with magnetospheres, plasma torii, solar wind, ring rain, interplanetary dust | NIR/MIR resolved spectroscopy: Jupiter/Saturn polar regions, Uranus/Neptune globally. | Diffraction-limited NIR imaging for short-term context, spatially-resolved spectroscopy for temperatures and abundances. | 4 | 5 h | 5 | 1 |
| Understand the structure and evolution of planetary rings, derive insights to planet formation processes, and constrain the interior state of giant planets [1,2,4,7,8,10] | Measure evolution of dynamical properties of rings and moons, and radial/asimuthal particle properties | High-resolution images of ring systems, covering 180°+ of ring orbital longitude | Diffraction-limited NIR imaging within CH4 absorption bands | 4 | 10 h | 0.5 | 1-4 |
| Understand the evolution of surface volatiles on icy bodies, including Ocean Worlds. [4,6,10] | Understand changes in H2O ice state, and in the abundance/spatial distribution of irradiation-produced volatiles | Spatially-resolved NIR spectroscopy of icy Galilean and saturnian satellites | Diffraction limited spectroscopy in 0.7-2.4 µm range at R~4000 | 12 | 2-80 d | 0.5 | 4 |
| | Measure seasonal changes in volatile CO2 | Unresolved NIR spectroscopy of uranian satellites | Point-source spectroscopy in K at R~4000 | 5 | 1-10 d | 0.5 | 4 |
| | Measure migration of hypervolatiles like CO, N2, CH4, including surface/atmosphere interactions | Spatially-resolved NIR spectroscopy of Triton, Pluto/Charon, TNO dwarf planets | Diffraction limited spectroscopy in 0.7-2.4 µm range at R~4000 | 10 | 0-5 d | 0.5 | 4 |

**Figure 4.** Traceability of science goals to observables to instrument and cadence requirements. Numbers in first column refer to Table S.1 in *Vision and Voyages for Planetary Science in the Decade 2013-2022* (p.19). "Mosaics" here are arrays of regularly-spaced pointings in time rather than space. Each pointing has a specific observation duration, and pointings are spaced over a span ranging from < 1 hr to many days.